\documentclass[%
 reprint,
 superscriptaddress,
 amsmath,amssymb,
 aps,
prl,
]{revtex4-1}

\newcommand{\tabincell}[2]{\begin{tabular}{@{}#1@{}}#2\end{tabular}}
\usepackage{graphicx}
\usepackage{dcolumn}
\usepackage{bm}
\usepackage{colortbl}
\usepackage{float}
\usepackage{hyperref}
\usepackage[table]{xcolor}
\usepackage{makecell}
\usepackage[mathscr]{euscript}
\usepackage[table]{xcolor}
\usepackage{multirow}
\usepackage{braket}
\usepackage[pagewise]{lineno}
\usepackage{enumitem}

\begin{document}

\title{The anti-symmetric and anisotropic symmetric exchange interactions \\
between electric dipoles in hafnia}

 \author{Longju Yu}
 \affiliation{International Center for Computational Method and Software, College of Physics, Jilin University, Changchun 130012, China}
\author{Hong Jian Zhao}
 \email{physzhaohj@jlu.edu.cn}
 \affiliation{International Center for Computational Method and Software, College of Physics, Jilin University, Changchun 130012, China}
 \affiliation{Key Laboratory of Physics and Technology for Advanced Batteries (Ministry of Education), College of Physics, Jilin University, Changchun 130012, China}
\affiliation{State Key Laboratory of Superhard Materials, College of Physics, Jilin University, Changchun 130012, China}
 \affiliation{International Center of Future Science, Jilin University, Changchun 130012, China}
 \author{Peng Chen}
\affiliation{Physics Department and Institute for Nanoscience and Engineering University of Arkansas, Fayetteville, Arkansas 72701, USA}
 \author{Laurent Bellaiche}
\affiliation{Physics Department and Institute for Nanoscience and Engineering University of Arkansas, Fayetteville, Arkansas 72701, USA}
\author{Yanming Ma}
\email{mym@jlu.edu.cn}
 \affiliation{International Center for Computational Method and Software, College of Physics, Jilin University, Changchun 130012, China}
 \affiliation{International Center of Future Science, Jilin University, Changchun 130012, China}
 \affiliation{State Key Laboratory of Superhard Materials, College of Physics, Jilin University, Changchun 130012, China}

\begin{abstract}
The anti-symmetric and anisotropic symmetric exchange interactions between two magnetic dipole moments -- responsible for  intriguing magnetic textures ({\it e.g.,} magnetic skyrmions) -- have been discovered since last century, while their electric analogues were either hidden for a long time or still not known. As a matter of fact, it is only recently that the anti-symmetric exchange interactions between electric dipoles was proved to exist (with materials hosting such an interaction being still rare) and the existence of anisotropic symmetric exchange interaction between electric dipoles remains to be revealed. Here, by symmetry analysis and first-principles calculations, we identify a candidate material in which our aforementioned exchange interactions between electric dipoles are perceptible. More precisely, we 
find that various phases of hafnia showcase non-collinear alignment of electric dipoles, which is interpreted by our phenomenological theories. This gives evidence that hafnia simultaneously accommodates anti-symmetric and anisotropic symmetric exchange interactions between electric dipoles. Our findings can hopefully deepen the current knowledge of electromagnetism in ferroelectrics, magnets and multiferroics, and have a potential to guide the discovery of novel states of matter ({\it e.g.}, electric skyrmions) in hafnia and related materials. 
\end{abstract}

\maketitle

\noindent
\textit{Introduction.--}In the last century, the profound exchange couplings between magnetic dipole moments -- namely, the magnetic anti-symmetric exchange interactions (also termed as magnetic Dzyaloshinskii-Moriya interaction, mDMI) and the magnetic anisotropic symmetric exchange interaction (mASEI) -- were derived with the origin attributed to spin-orbit interaction (see {\it e.g.}, Refs.~\cite{dzyaloshinsky1957thermodynamic,dzyaloshinsky1958thermodynamic,moriya1960anisotropic}). The mDMI and mASEI are physical underpinnings for many intriguing non-collinear magnetic textures ({\it e.g.}, magnetic vortices~\cite{mvortices,mvortices2}, skyrmions~\cite{wang2018ferroelectrically,foster2019two,amoroso2020spontaneous,legrand2020room,yu2018transformation,yangdmi2023} and merons~\cite{yu2018transformation,gao2019creation,bera2019theory,yangdmi2023}) that are promising for novel device applications in information technology~\cite{fert2017magnetic,bogdanov2020physical,tokura2020magnetic}. Strikingly, {\it  electric} vortices, skyrmions and merons have also been observed and/or predicted in ferroelectric nanostructures such as Pb(Zr,Ti)O$_3$ nanodisks, nanorods and thin films~\cite{naumov2004unusual,pztfilm,nahas2020}, BaTiO$_3$-SrTiO$_3$ nanocomposites~\cite{nahas2015discovery}, PbTiO$_3$ thin films~\cite{wang2020polar}, PbTiO$_3$ nanodomains~\cite{pereira2019theoretical} and SrTiO$_3$/PbTiO$_3$ superlattices~\cite{hong2017stability,yadav2016observation,das2019observation,das2021local,han2022high}.
The mechanisms for these non-collinear dipolar textures were mostly ascribed to the depolarizing field or the Bloch-like domain walls, instead of the exchange interactions between electric dipoles~\cite{naumov2004unusual,nahas2015discovery,pereira2019theoretical,wang2020polar,hong2017stability,yadav2016observation,das2019observation,das2021local,han2022high}. To understand the non-collinear ferroelectricity in bulk materials ({\it i.e.}, with no depolarizing field or domain wall)~\cite{khalyavin2020emergent,lin2019frustrated,varignon2016electric,zhao2014creating,yang2012revisiting,yang2012epitaxial,belik2006biino3},
the exchange interactions between electric dipoles were recently revisited, yielding the disclosure of the electric Dzyaloshinskii-Moriya interaction (eDMI)~\cite{zhao2021dzyaloshinskii,hlinka2020,chen2022microscopic} -- which may also be responsible for the recently observed and so-called double-$\mathbf{Q}$-modulated structure ~\cite{rusu2022ferroelectric}. Even so, materials that are known to host eDMI remain rare in nature. Besides, whether there exists electric anisotropic symmetric exchange interaction (eASEI) is still currently elusive.

Here, via symmetry analysis and first-principles calculations, we identify hafnia (HfO$_2$) material as an ideal candidate that accommodates the exchange interaction (eDMI and eASEI) between electric dipoles. We show that HfO$_2$ has various polymorphisms ({\it i.e.}, $P2_1/c$, $Pmn2_1$, $Pca2_1$ and $Pbca$ phases) demonstrating non-collinear alignment of electric dipoles. The non-collinear dipole patterns (NCDP) herein are interpreted by our phenomenological theories, further revealing that (i) eDMI and eASEI are simultaneously hosted by HfO$_2$, and that (ii) both eDMI and eASEI stem from the structural distortions associated with O sublattice. \\

\begin{figure*}[htb]
\includegraphics[width=0.85\linewidth]{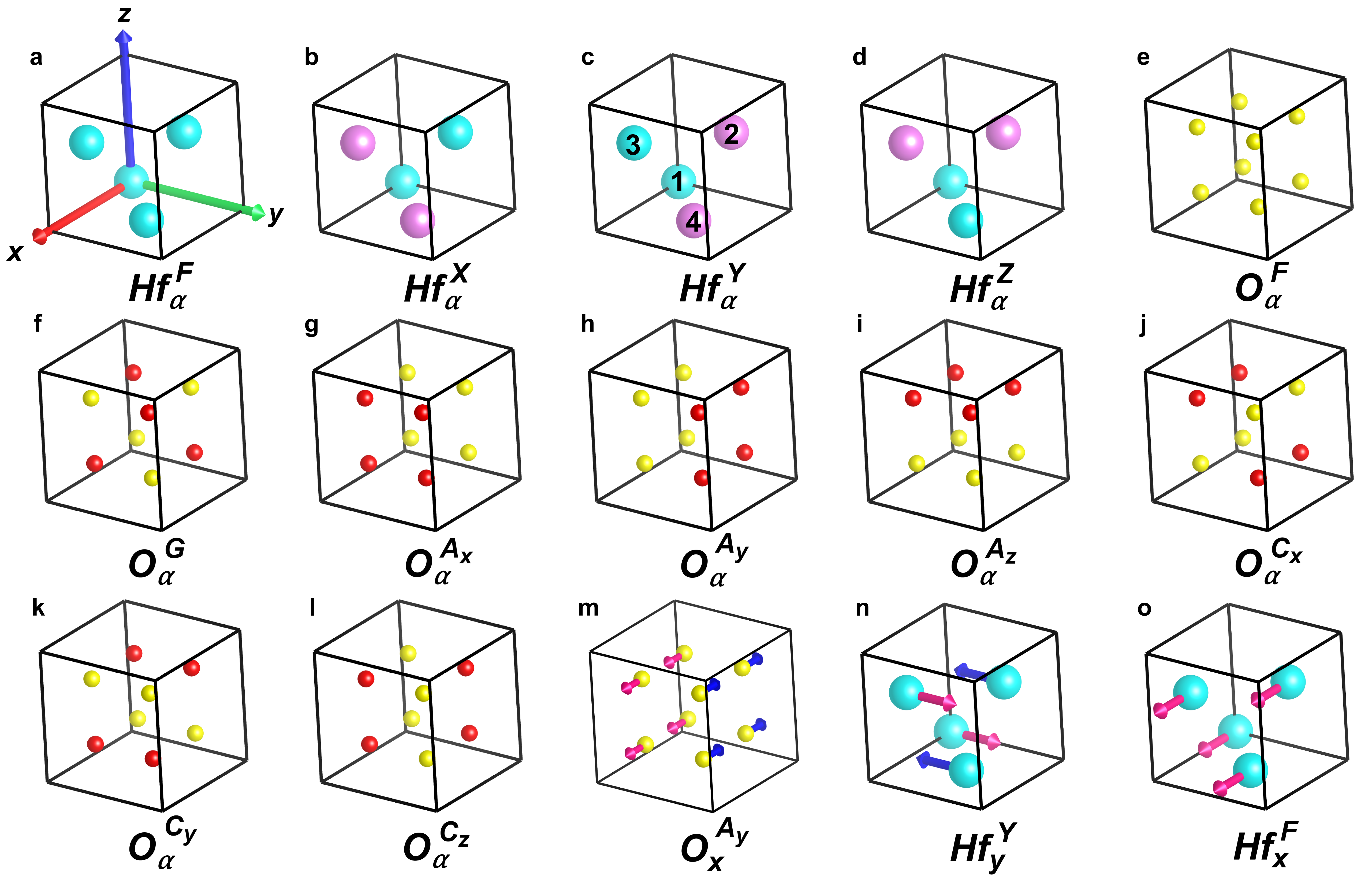}
\caption{\label{fig:modulations}Sketches of Hf and O motions in $\rm HfO_{2}$ with respect to the conventional cell of the $Fm\bar{3}m$ phase. In panels (a)-(d), the cyan and pink spheres indicate that the Hf's displacements centered on the corresponding spheres are along $+\alpha$ and $-\alpha$ directions, respectively. For displaying clarity, we do not show the periodic image of the Hf atoms in each cell. Panels (e)-(l) sketch various configurations associated with the O sublattice. In these sketches, the yellow and red spheres denote that the motions centered on the corresponding O sites are along $+\alpha$ or $-\alpha$ directions, respectively. Panels (m), (n) and (o) display three examples of atomic motions with the configurations defined by panels (h), (c) and (a) [the corresponding $\alpha$ directions being $x$, $y$ and $x$], where pink and blue arrows indicate two motions of opposite directions.}
\end{figure*}

\noindent
\textit{The NCDP in HfO$_2$'s structural phases.--}Experimentally, HfO$_2$ was found to be polymorphic, with a variety of structural phases such as $Fm\bar{3}m$~\cite{wang1992hafnia}, $P4_2/nmc$~\cite{curtis1954some}, $Pbca$~\cite{Ohtaka1991SynthesisAX}, $Pnma$~\cite{liu1980new}, $Pbcm$~\cite{pathak2020structural}, $Pca2_1$~\cite{xu2021kinetically}, and $P2_1/c$~\cite{hann1985monoclinic}. Recent works by first-principles simulations also highlight the possibility of the polar $Pmn2_1$ phase of HfO$_2$ (see, {\it e.g.}, Refs.~\cite{PhysRevLett.125.257603,PhysRevB.90.064111,Batra2016StabilizationOM}). 
Of particular interest are the $P2_1/c$, $Pmn2_1$, $Pca2_1$ and $Pbca$ phases, since (i) they exhibit NCDP~\footnote{For simplicity, the electric dipoles discussed in the present work are centered on Hf sites and denoted by the motions (see Fig.~\ref{fig:modulations}) of Hf ions.} that imply the existence of eDMI and eASEI, and (ii) such NCDP can be captured by our developed phenomenological theories (see Table~\ref{tab:Trilinearcouplings}).

Before extracting the NCDP in $P2_1/c$, $Pmn2_1$, $Pca2_1$ and $Pbca$ phases, we analyze the possible structural distortions accommodated by the high-symmetric $Fm\bar{3}m$ phase of HfO$_2$. For simplicity, we select the conventional cell of $Fm\bar{3}m$ HfO$_2$ -- containing four formula units -- as our reference structure~\footnote{As will be shown below, our phenomenological theories based on the conventional cell of $Fm\bar{3}m$ HfO$_3$ can well describe the NCDP in $P2_1/c$, $Pmn2_1$, $Pca2_1$ and $Pbca$ phases.
Using a larger cell, although captures more abundant structural distortions and NCDP, will significantly increase the difficulties for our symmetry analysis.}. In such a conventional cell, there are four Hf ions and eight O ions, whose motions~\footnote{The motion is referred to as the displacement of a ion with respect to its original position in the reference $Fm\bar{3}m$ phase.} constitute thirty-six structural order parameters. In the present work, the order parameters contributed by Hf sublattice are labelled by $Hf^{U}_\alpha$ ($U=F,X,Y,Z$; $\alpha=x,y,z$). Here, the subscript $\alpha$ and superscript $U$ denote that the atomic displacement for each Hf ion is either parallel or antiparallel to $\alpha$ direction, depending on the configuration labelled by $U$ as defined in Figs.~\ref{fig:modulations}(a)-\ref{fig:modulations}(d). For example, Fig.~\ref{fig:modulations}(a) depicts the $Hf^F_\alpha$ configuration whose atomic motions are homogeneous with respect to the four Hf ions in the cell. When endowing the $Hf^F_\alpha$ configuration with Hf's displacement along $x$ direction, we arrive at the $Hf^F_x$ order parameter sketched in Fig.~\ref{fig:modulations}(o). Another example is the $Hf^Y_\alpha$ configuration and the $Hf^Y_y$ order parameter, shown in Figs.~\ref{fig:modulations}(c) and~\ref{fig:modulations}(n). In the $Hf^Y_\alpha$ configuration, the displacements for Hf ions numbered by 1 and 3 (respectively, 2 and 4) are along $+\alpha$ (respectively, $-\alpha$) orientations; the atomic motions along $y$ direction associated with $Hf^Y_\alpha$ configuration is thus marked as $Hf^Y_y$. Similarly, we can define the other order parameters associated with Hf sublattice [see Figs.~\ref{fig:modulations}(b) and~\ref{fig:modulations}(d)] and those contributed by the O sublattice [see Figs.~\ref{fig:modulations}(e)-\ref{fig:modulations}(m)] in a self-explanatory manner. Following this convention, we have identified thirty-six order parameters for HfO$_2$ [see Section I of the Supplementary Material (SM) for details].

\begin{table}[h]\footnotesize
\caption{\label{tab:Trilinearcouplings}Trilinear couplings resulting in NCDP for various phases of HfO$_2$. Here, the definitions of the notations ({\it e.g.,} $Hf^{Y}_{y}$ and $ O^{A_{x}}_{y}$) are indicated in Fig.~\ref{fig:modulations}. The trilinear coupling associated with a specific phase of HfO$_2$ is shown in the parentheses after the space group of that phase.}
\begin{ruledtabular}
\renewcommand{\arraystretch}{1.5}
\begin{tabular}{cc}

\small Hamiltonian&\small Phases\\ \hline
\multirow{3}*{ \tabincell{c}{ $H_1 \propto {Hf}^{Y}_{y} { Hf}^{Z}_{x} { O}^{A_{x}}_{y} + { Hf}^{Y}_{x} { Hf}^{Z}_{z} { O}^{A_{x}}_{z} $\\$\hspace{1em} +{Hf}^{X}_{x} {Hf}^{Z}_{y} {O}^{A_{y}}_{x} + {Hf}^{X}_{y} {Hf}^{Z}_{z} { O}^{A_{y}}_{z} $\\$\hspace{1em} +{ Hf}^{X}_{x} { Hf}^{Y}_{z} {O}^{A_{z}}_{x} + 
{ Hf}^{X}_{z} { Hf}^{Y}_{y} { O}^{A_{z}}_{y}$}}

&$P2_{1}/c$\ (${Hf}^{X}_{x} { Hf}^{Z}_{y} { O}^{A_{y}}_{x}$)\\

~&$Pca2_{1}$\ (${Hf}^{Y}_{y} { Hf}^{Z}_{x} { O}^{A_{x}}_{y}$)\\

&$Pbca$\ (${Hf}^{X}_{z} { Hf}^{Y}_{y} { O}^{A_{z}}_{y}$)\\
\hline

\multirow{3}*{\tabincell{c}{$H_2 \propto  { Hf}^{F}_{y} { Hf}^{X}_{x} { O}^{A_{x}}_{y} +  { Hf}^{F}_{z} { Hf}^{X}_{x} { O}^{A_{x}}_{z} $\\$\hspace{1em} +{Hf}^{F}_{x} { Hf}^{Y}_{y} { O}^{A_{y}}_{x} + { Hf}^{F}_{z} { Hf}^{Y}_{y} { O}^{A_{y}}_{z}  $\\$\hspace{1em} +{ Hf}^{F}_{x} { Hf}^{Z}_{z} { O}^{A_{z}}_{x} + { Hf}^{F}_{y} { Hf}^{Z}_{z} { O}^{A_{z}}_{y}$}}

&\multirow{2}*{$Pmn2_{1}$\ \tabincell{c}{(${Hf}^{F}_{x} { Hf}^{Y}_{y} { O}^{A_{y}}_{x}$) \\(${Hf}^{F}_{z} { Hf}^{Y}_{y} { O}^{A_{y}}_{z}$)  }}\\ \\

~&$Pca2_{1}$\ (${Hf}^{F}_{z} { Hf}^{Y}_{y} { O}^{A_{y}}_{z}$)\\

\hline

\multirow{3}*{\tabincell{c}{$H_3 \propto  { Hf}^{F}_{x} { Hf}^{X}_{y} { O}^{A_{x}}_{y} +{ Hf}^{F}_{x} { Hf}^{X}_{z} { O}^{A_{x}}_{z} $\\$\hspace{1em} +{ Hf}^{F}_{y} { Hf}^{Y}_{x} { O}^{A_{y}}_{x} + { Hf}^{F}_{y} { Hf}^{Y}_{z} { O}^{A_{y}}_{z} $\\$\hspace{1em} +{ Hf}^{F}_{z} { Hf}^{Z}_{x} { O}^{A_{z}}_{x} + {Hf}^{F}_{z} {Hf}^{Z}_{y} { O}^{A_{z}}_{y}$}}

&\multirow{3}*{$Pca2_{1}$\ (${Hf}^{F}_{z} { Hf}^{Z}_{x} { O}^{A_{z}}_{x}$)}\\ \\
\\
\hline

\multirow{3}*{\tabincell{c}{$H_4 \propto { Hf}^{Y}_{x} { Hf}^{Z}_{y} { O}^{A_{x}}_{y} + { Hf}^{Y}_{z} { Hf}^{Z}_{x} { O}^{A_{x}}_{z} $\\$\hspace{1em} +{ Hf}^{X}_{y} { Hf}^{Z}_{x} { O}^{A_{y}}_{x} +{ Hf}^{X}_{z} { Hf}^{Z}_{y} { O}^{A_{y}}_{z} $\\$\hspace{1em} +{ Hf}^{X}_{z} { Hf}^{Y}_{x} { O}^{A_{z}}_{x} + {Hf}^{X}_{y} { Hf}^{Y}_{z} { O}^{A_{z}}_{y}$}}

&\multirow{3}*{$P2_{1}/c$\ (${Hf}^{X}_{z} { Hf}^{Z}_{y} { O}^{A_{y}}_{z}$)}  \\ \\

\\
\end{tabular}
\end{ruledtabular}
\end{table}

\begin{figure}[htb]
\includegraphics[width=1.02\linewidth]{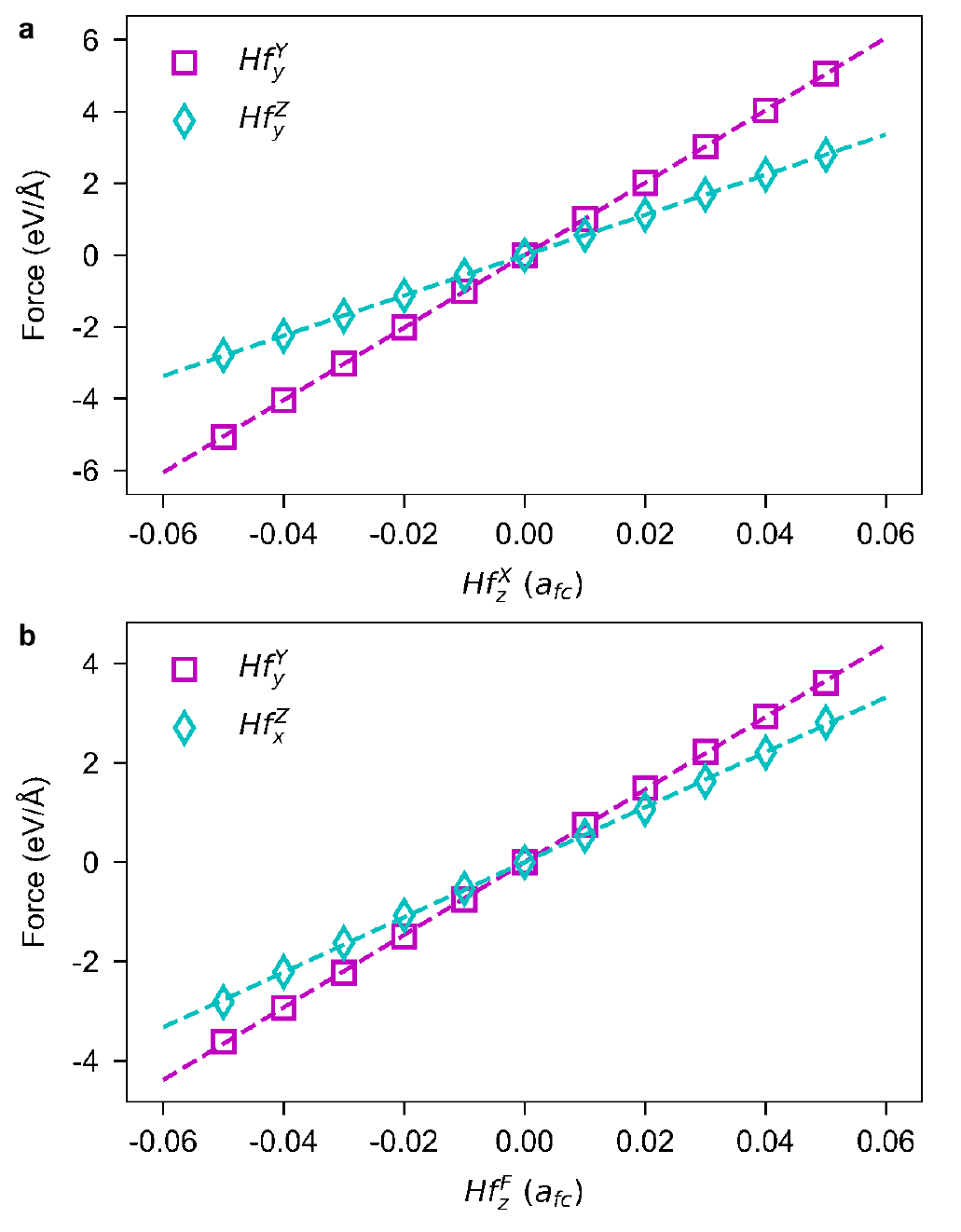}
\caption{\label{fig:Pca21force}Forces on Hf sublattice in HfO$_2$ as a function of 
$Hf^{X}_{z}$ (a) and $Hf^{F}_{z}$ (b) distortions. Purple square in panel (a): forces associated with $Hf^{Y}_{y}$ mode ($O^{A_{z}}_{y}$ being fixed). Cyan diamond in panel (a): forces associated with $Hf^{Z}_{y}$ mode ($O^{A_{y}}_{z}$ being fixed). Purple square in panel (b): forces associated with $Hf^{Y}_{y}$ mode ($O^{A_{y}}_{z}$ being fixed). Cyan diamond in panel (b): forces associated with $Hf^{Z}_{x}$ mode ($O^{A_{z}}_{x}$ being fixed). The dash lines in panels (a) and (b) display the linear fitting results, corresponding to $Hf^{X}_{z} Hf^{Y}_{y} O^{A_z}_y$, $Hf^{X}_{z} Hf^{Z}_{y} O^{A_y}_z$, $Hf^{F}_{z} Hf^{Y}_{y} O^{A_y}_z$ and $Hf^{F}_{z} Hf^{Z}_{x} O^{A_z}_x$ couplings, respectively. }
\end{figure}

Starting from these thirty-six order parameters, we construct phenomenological theories that describe  NCDP in HfO$_2$. We notice that the combination of $Hf^U_\alpha$ and $Hf^V_\beta$ order parameters naturally yields NCDP, when $U \neq V$ and $\alpha \neq \beta$. By symmetry arguments, $Hf^U_\alpha$ and $Hf^V_\beta$ are possibly coexisting via the $Hf^U_\alpha Hf^V_\beta O^W_\gamma$ trilinear coupling, that mediated by the $O^W_\gamma$ structural order parameter. As shown in Section I of the SM, we have derived four effective Hamiltonians ($H_1$, $H_2$, $H_3$, and $H_4$) involving trilinear couplings of our aforementioned kind, summarized in Table~\ref{tab:Trilinearcouplings}. We can verify the existence of the couplings in $H_l$ ($l=1-4$) by first-principles numerical calculations, using the following strategy (see Section II of the SM for details):
to verify the $Hf^U_\alpha Hf^V_\beta O^W_\gamma$ coupling, we (i) start from the $Fm\bar{3}m$ phase and impose a structural distortion according to $O^W_\gamma$ with fixed amplitude, (ii) displace Hf ions following $Hf^V_\beta$ mode with varying magnitude, and (iii) measure the first-principles-calculated forces acting on the Hf sublattice and associated with the $Hf^U_\alpha$ mode. The linear relationship between these forces (related to $Hf^U_\alpha$) and the distortion amplitudes (of $Hf^V_\beta$) will corroborate the existence of the $Hf^U_\alpha Hf^V_\beta O^W_\gamma$ coupling. Fig.~\ref{fig:Pca21force} indeed numerically confirms the existence of several selective trilinear couplings, namely, $Hf^{X}_{z} Hf^{Y}_{y} O^{A_z}_y$, $Hf^{X}_{z} Hf^{Z}_{y} O^{A_y}_z$, $Hf^{F}_{z} Hf^{Y}_{y} O^{A_y}_z$ and $Hf^{F}_{z} Hf^{Z}_{x} O^{A_z}_x$. Interestingly, our derived $Hf^F_z Hf^Z_x O^{A_z}_x$ coupling coincides with the trilinear coupling that was claimed to drive the ferroelectricity of $Pca2_1$ HfO$_2$ (see Ref.~\cite{HfO2picozzi}).

\begin{figure}[htb]
\includegraphics[width=1.0\linewidth]{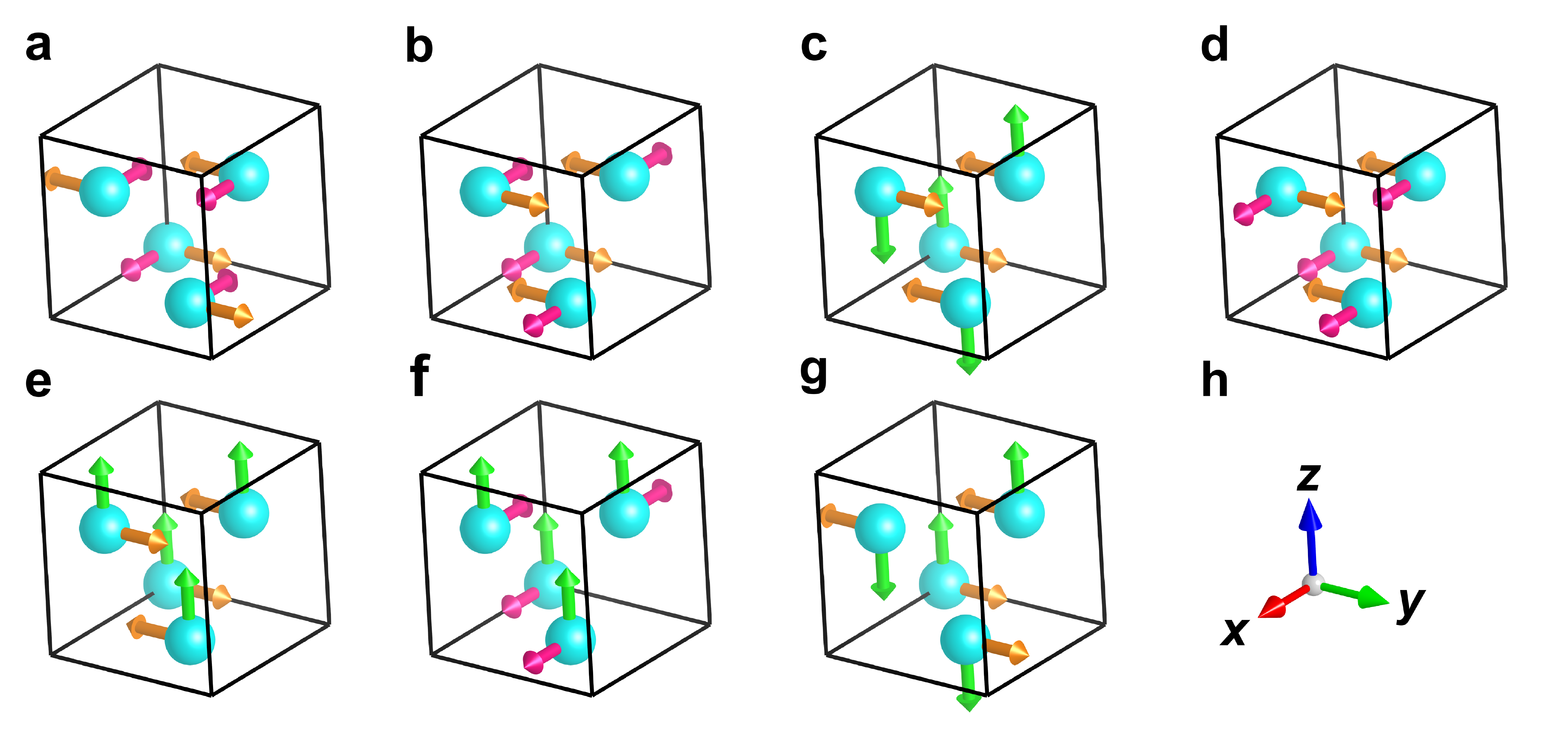}
\caption{\label{fig:dualcoupling}Sketches of atomic motions associated with various trilinear couplings. The ${Hf}^{X}_{x} {Hf}^{Z}_{y} { O}^{A_{y}}_{x}$, ${Hf}^{Y}_{y} {Hf}^{Z}_{x} { O}^{A_{x}}_{y}$, ${Hf}^{X}_{z} {Hf}^{Y}_{y} { O}^{A_{z}}_{y}$, ${Hf}^{F}_{x} {Hf}^{Y}_{y} { O}^{A_{y}}_{x}$,  ${Hf}^{F}_{z} {Hf}^{Y}_{y} { O}^{A_{y}}_{z}$, ${Hf}^{F}_{z} {Hf}^{Z}_{x} { O}^{A_{z}}_{x}$ and ${Hf}^{X}_{z} {Hf}^{Z}_{y} { O}^{A_{y}}_{z}$ couplings are depicted in panels (a)-(g). Our coordinate system is indicated in panel (h).  For instance, panel (a) denotes the coexistence of $Hf^{X}_{x}$ and $Hf^{Z}_{y}$ mediated by $O^{A_{y}}_{x}$. The pink, orange and green arrows denote the 
motions of Hf being parallel or anti-parallel to the $x$, $y$ and $z$ directions, respectively. For clarity, we do not show the O motions.}
\end{figure}

We now extract the NCDP in HfO$_2$ and link the NCDP with our phenomenological theories. In Section III of the SM, we analyze the structural distortions in $P2_1/c$, $Pmn2_1$, $Pca2_1$ and $Pbca$ phases of HfO$_2$. In $P2_1/c$ phase, the $Hf^{X}_{x} Hf^{Z}_{y} O^{A_{y}}_{x}$ and $Hf^{X}_{z} Hf^{Z}_{y} O^{A_{y}}_{z}$ trilinear couplings -- see Table~\ref{tab:Trilinearcouplings} -- imply the $(Hf^{X}_x,Hf^{Z}_y)$ and $(Hf^{X}_z,Hf^{Z}_y)$ combinations, respectively. As sketched in Figs.~\ref{fig:dualcoupling}(a) and~\ref{fig:dualcoupling}(g), the $(Hf^{X}_x,Hf^{Z}_y)$ and $(Hf^{X}_z,Hf^{Z}_y)$ combinations yield NCDP. The NCDP in $Pca2_1$ phase seems more complicated. To be specific, the $(Hf^{Y}_y,Hf^{Z}_x)$, $(Hf^{Y}_y,Hf^{F}_z)$ and $(Hf^{F}_z,Hf^{Z}_x)$ combinations [see Figs.~\ref{fig:dualcoupling}(b),~\ref{fig:dualcoupling}(e), and~\ref{fig:dualcoupling}(f)] can yield the NCDP. These combinations come from the $Hf^{Y}_y Hf^{Z}_x O^{A_{x}}_{y}$, $Hf^{F}_z Hf^{Y}_y O^{A_{y}}_{z}$ and $Hf^{F}_z Hf^{Z}_x O^{A_{z}}_{x}$ trilinear couplings. Furthermore, the $Hf^{F}_x Hf^{Y}_y O^{A_{y}}_{x}$ and $Hf^{F}_z Hf^{Y}_y O^{A_{y}}_{z}$ couplings lead to the NCDP in $Pmn2_1$ phase [via $(Hf^{F}_x,Hf^{Y}_y)$ and $(Hf^{F}_z,Hf^{Y}_y)$ combinations, see Figs.~\ref{fig:dualcoupling}(d) and~\ref{fig:dualcoupling}(e)], while the $Hf^{X}_z Hf^{Y}_y O^{A_z}_y$ coupling gives rise to the NCDP in the $Pbca$ phase [via $(Hf^{X}_z, Hf^{Y}_y)$ combination, see Fig.~\ref{fig:dualcoupling}(c)].
Our aforementioned analysis thus emphasizes the importance of the $Hf^U_\alpha Hf^V_\beta O^W_\gamma$-type of trilinear couplings ($U \neq V$, $\alpha \neq \beta$) towards the NCDP in HfO$_2$'s structural phases. Here, the central structural distortion is $O^W_\gamma$ and is thus contributed by the O sublattice, mediating then the  interaction between $Hf^U_\alpha$ and $Hf^V_\beta$ distortions. In other words, the $O^W_\gamma$-type distortion is the structural origin of the NCDP in HfO$_2$.\\

\noindent
\textit{The anti-symmetric exchange interaction.--}The correlation between $Hf^U_\alpha Hf^V_\beta O^W_\gamma$ couplings ($U\neq V$, $\alpha \neq \beta$) and NCDP opens a door to reveal the anti-symmetric exchange interactions (eDMI) of dipoles in HfO$_2$ oxide. 
To begin with, we recall that the magnetic exchange interaction is given by \cite{moriya1960anisotropic,amoroso2020spontaneous}
\begin{equation}\label{magexch}
\mathcal{H}=\sum_{i\neq j,\alpha,\beta} J_{ij,\alpha\beta}m_{i,\alpha}m_{j,\beta},
\end{equation}
where (i) $m_{i,\alpha}$ and $m_{j,\beta}$ ($\alpha,\beta=x,y,z$) are $\alpha$- and $\beta$-component of magnetic dipole moments centered on the $i_\mathrm{th}$ and $j_\mathrm{th}$ ions, respectively, and (ii) $J_{ij,\alpha\beta}$ characterizes the strength of coupling between $m_{i,\alpha}$ and $m_{j,\beta}$. Eq.~(\ref{magexch}) implies that the electric exchange interaction between $u_{i,\alpha}$ and $u_{j,\beta}$ dipoles (if it exists) can be written as 
\begin{equation}\label{elecexch}
H=\sum_{i\neq j,\alpha,\beta} J^\prime_{ij,\alpha\beta}u_{i,\alpha}u_{j,\beta}.
\end{equation}
Here, $u_{i,\alpha}$ and $u_{j,\beta}$ are atomic displacements, depicting the electric dipoles centered on $i_\mathrm{th}$ and $j_\mathrm{th}$ ions.

To evaluate $J^\prime_{ij,\alpha\beta}$ in HfO$_2$, we start from the $Fm\bar{3}m$ phase and work with a big supercell made of $N$ conventional cells (see Section IV of the SM for details). 
Such a supercell contains $4N$ Hf ions with their atomic coordinates given by $\mathbf{R}_m+\mathbf{r}_\tau$, where $\mathbf{R}_m$ locates the $m_\mathrm{th}$ conventional cell ($m=1,2,...,N$) and $\mathbf{r}_\tau$ is the coordinate of Hf inside the $m_\mathrm{th}$ cell ($\tau=1,2,3,4$, see Fig. S6 of the SM). Every Hf ion in the supercell can displace along the $\alpha$ direction ($\alpha=x,y,z$) with respect to $\mathbf{R}_m+\mathbf{r}_\tau$, creating a dipole $u_{m,\tau,\alpha}$~\footnote{Comparing $u_{m,\tau,\alpha}$ with $u_{i,\alpha}$ implies that $m,\tau \equiv i$}. 
We expand the $Hf^F_\alpha$, $Hf^X_\alpha$, $Hf^Y_\alpha$, and $Hf^Z_\alpha$ order parameters by the $u_{m,\tau,\alpha}$ basis, and insert these expansions into our derived $H_l$ ($l=1-4$), as demonstrated in Eqs. (S1)-(S6) of the SM. This yields the effective Hamiltonian as 
\begin{equation}\label{elecexchhfo2}
H_l=\sum_{m,m^\prime,\tau,\kappa,\alpha,\beta} J^\prime_{m\tau m^\prime \kappa,\alpha\beta} u_{m,\tau,\alpha}u_{m^\prime,\kappa,\beta},
\end{equation}
where $J^\prime_{m\tau m^\prime \kappa,\alpha\beta}$ -- a function of $O^W_\gamma$, $m$, $m^\prime$, $\kappa$, $\tau$, $\alpha$, and $\beta$ -- characterizes the coupling between $u_{m,\tau,\alpha}$ and $u_{m^\prime,\kappa,\beta}$ dipoles. As a result, the $J^\prime_{m\tau m^\prime \kappa,\alpha\beta}$ interaction associated with the $H_l$ Hamiltonian can be extracted via
\begin{equation}\label{Jelements}
J^\prime_{m\tau m^\prime \kappa,\alpha\beta}=\frac{\partial^2 H_l}{\partial u_{m,\tau,\alpha} \partial u_{m^\prime,\kappa,\beta}}.
\end{equation}
and the strength of the eDMI is evaluated  by~\footnote{The eDMI between $\mathbf{u}_{i}\equiv(u_{i,x},u_{i,y},u_{i,z})$ and $\mathbf{u}_{j}\equiv(u_{j,x},u_{j,y},u_{j,z})$ dipoles is defined as $\mathbf{D}_{ij}^\prime\cdot(\mathbf{u}_{i}\times\mathbf{u}_{i})$ with $\mathbf{D}_{ij}^\prime\equiv(D_{ij,x}^\prime,D_{ij,y}^\prime,D_{ij,z}^\prime)$ being the eDMI vector (see Refs.~\cite{zhao2021dzyaloshinskii,chen2022microscopic}). Expanding $\mathbf{D}_{ij}^\prime\cdot(\mathbf{u}_{i}\times\mathbf{u}_{i})$ results in $D^\prime_{ij,x}(u_{i,y}u_{j,z}-u_{i,z}u_{j,y})+D^\prime_{ij,y}(u_{i,z}u_{j,x}-u_{i,x}u_{j,z})+D^\prime_{ij,z}(u_{i,x}u_{j,y}-u_{i,y}u_{j,x})$. By $i\rightarrow m\tau$ and $j \rightarrow m^\prime\kappa$ replacements, such an expansion together with Eqs.~(\ref{elecexch})-(\ref{Jelements}) yield Eq.~(\ref{Aelements}), where $D^\prime_{m\tau m^\prime\kappa,x}=A^\prime_{m\tau m^\prime\kappa,yz}$, $D^\prime_{m\tau m^\prime\kappa,y}=A^\prime_{m\tau m^\prime\kappa,zx}$, and $D^\prime_{m\tau m^\prime\kappa,z}=A^\prime_{m\tau m^\prime\kappa,xy}$.}
\begin{equation}\label{Aelements}
A^\prime_{m\tau m^\prime \kappa,\alpha\beta}=\frac{1}{2}(J^\prime_{m\tau m^\prime \kappa,\alpha\beta}-J^{\prime}_{m\tau m^\prime \kappa,\beta\alpha}).
\end{equation}

In the following, we focus on the eDMI associated with two neighbored Hf ions which belong to the same conventional cell ({\it e.g.}, $m=m^\prime$, $\tau \neq \kappa$). Our detailed derivation process is shown in Section IV of the SM; here, we omit the cell labels $m$ amd $m^\prime$. With respect to each $H_l$ effective Hamiltonian, the $A^\prime_{\tau \kappa,\alpha\beta}$ components -- for the eDMI between Hf$_\tau$ and Hf$_\kappa$ pair ($\tau,\kappa=1,2,3,4$) -- form a $3\times 3$ anti-symmetric matrix. As shown in Tables S2, S4, S6 and S8 of the SM, the $A^\prime_{\tau  \kappa,\alpha\beta}$ component is determined by the $O^W_\gamma$-type distortion contributed by the O sublattice. For example, we examine the interaction involving Hf$_1$ and Hf$_2$ ions, where $\mathbf{r}_\tau \equiv \mathbf{r}_1=0$ and $\mathbf{r}_\kappa \equiv \mathbf{r}_2=0\mathbf{a}+\frac{1}{2}\mathbf{b}+\frac{1}{2}\mathbf{c}$ ($\mathbf{a}$, $\mathbf{b}$ and $\mathbf{c}$ being the lattice vectors of $Fm\bar{3}m$'s conventional cell). The $H_1$ Hamiltonian suggests that $A^\prime_{12,xy}\propto -O^{A_y}_x$ [see Eq.~(S9) and Table S2 of the SM].\\

\noindent
\textit{The anisotropic symmetric exchange interaction.--}We move on to explore the eASEI that may be hosted by HfO$_2$. Compared with the mAESI (see {\it e.g.}, Refs.~\cite{moriya1960anisotropic,amoroso2020spontaneous}), the eASEI between $\mathbf{u}_{i}\equiv(u_{i,x},u_{i,y},u_{i,z})$ and $\mathbf{u}_{j}\equiv(u_{j,x},u_{j,y},u_{j,z})$ dipoles (if it exists) can be defined by $\sum_{\alpha \beta} S^\prime_{ij,\alpha\beta}u_{i,\alpha}u_{j,\beta}$, where $\alpha,\beta=x,y,z$, $S^\prime_{ij,\alpha\beta}=S^\prime_{ij,\beta\alpha}$, and $S^\prime_{ij,xx}+S^\prime_{ij,yy}+S^\prime_{ij,zz}=0$. Working with Eqs.~(\ref{elecexch}) and~(\ref{Jelements}), the strength of the eASEI between $u_{m,\tau,\alpha}$ and $u_{m^\prime,\kappa,\beta}$ is extracted by
\begin{equation}\label{Selements}
\begin{split}
S^\prime_{m\tau m^\prime \kappa,\alpha\beta}  =
\frac{1}{2}(J^\prime_{m\tau m^\prime \kappa,\alpha\beta}+J^{\prime}_{m\tau m^\prime \kappa,\beta\alpha}) \\
   -\frac{1}{3}\delta_{\alpha,\beta}(J^\prime_{m\tau m^\prime \kappa,xx}+J^\prime_{m\tau m^\prime \kappa,yy}+J^\prime_{m\tau m^\prime \kappa,zz}),
\end{split}
\end{equation}
where $\delta_{\alpha,\beta}=1$ for $\alpha=\beta$ and $\delta_{\alpha,\beta}=0$ otherwise. The $\alpha\beta$-components of $S^\prime_{m\tau m^\prime\kappa,\alpha\beta}$ form a $3\times3$ matrix that is symmetric and traceless.
In Section IV and Tables S3, S5, S7 and S9 of the SM, we calculate the eASEI between $\mathbf{u}_{m\tau}$ and $\mathbf{u}_{m\kappa}$ dipoles in HfO$_2$, following $H_l$ ($l=1-4$). The $S^\prime_{\tau  \kappa,\alpha\beta}$ (cell label $m$ being omitted) component is proportional to the $O^W_\gamma$-type distortion contributed by the O sublattice~\footnote{Even though $A^\prime_{\tau  \kappa,\alpha\beta}$ and $S^\prime_{\tau  \kappa,\alpha\beta}$ are both rooted in the $O^W_\gamma$-type distortion, their dependences on $O^W_\gamma$ are different. For example, our $H_1$ Hamiltonian indicates that $A^\prime_{12,xy} \propto -O^{A_y}_x$ and $S^\prime_{12,xy} \propto -O^{A_x}_y$ (see Tables S2 and S3 of SM). }.

So far, our discussion is based on $Hf^U_\alpha Hf^V_\beta O^W_\gamma$-type couplings ($U\neq V$, $\alpha \neq \beta$) -- as indicated in $H_1$, $H_2$, $H_3$, and $H_4$ -- that are linked with NCDP. As a by-product, we additionally obtain seven other effective Hamiltonians $H_l$ ($l=5-11$).  In contrast to $H_l$ ($l=1-4$) associated with NCDP, $H_l$ ($l=5-7$) and $H_l$ ($l=8-11$) are effective Hamiltonians with the types of $Hf^{U}_\alpha Hf^{U}_\beta O^{W}_\gamma$ ($\alpha \neq \beta$) and $Hf^{U}_\alpha Hf^{V}_\alpha O^{W}_\gamma$ ($U\neq V$), respectively, describing the collinear alignment of dipoles. As shown in Section IV of the SM, these Hamiltonians $H_l$ with $l=5-11$ yield the eASEI as well (see Tables S10-S16 in SM), with the structural origin being the $\omega^W_\gamma$-type distortion.

To complete our discussion on eASEI, we shall clarify the difference between eASEI and the interaction of local modes (in essence, electric dipoles). In ferroelectric theory, the interaction of local modes is written as $\mathscr{H}=\sum_{i\neq j,\alpha\beta}\mathscr{J}_{ij,\alpha\beta} \mu_{i,\alpha} \mu_{j,\beta}$~\cite{vanderbilt1995}, where $\mathscr{J}_{ij,\alpha\beta}$ characterizes the strength of the interaction, and $\mu_{i,\alpha}$ is the amplitude of the local mode centered on the $i_\mathrm{th}$ cell. Mathematically, the $\mathscr{H}$ interaction seems to resemble our proposed exchange interaction between electric dipoles [see Eq.~\ref{elecexch}]. From physical point of view, however, the two interactions are different. To be specific, the interaction between local models is a bilinear coupling between atomic motions ($\mathscr{J}_{ij,\alpha\beta}$ being independent of atomic motions), while the eASEI involves trilinear coupling among atomic motions--that is, $J^\prime_{m\tau m^\prime \kappa,\alpha\beta}$ depends on $O^W_\gamma$ structural distortion.\\

\noindent
\textit{Summary and outlook.--}In summary, we show that various structural phases of HfO$_2$ exhibit NCDP. By constructing phenomenological theories, we interpret the NCDP in HfO$_2$ and further reveal that NCDP are rooted in the exchange interactions (eDMI and eASEI) of electric dipoles. This implies a possible marriage between HfO$_2$-based oxides -- high-profile materials in semiconductor technology because of their compatibility with silicon~\cite{wei2018rhombohedral,schroeder2022fundamentals,xu2021kinetically,
nukala2021reversible,cheema2020enhanced,noheda2020key,lee2020scale,yun2022intrinsic,cheema2022ultrathin,cheema2022emergent,kang2022highly} -- and the topological textures of electric dipoles ({\it e.g.}, electric skyrmions), which are desired states of matter towards the creation of novel information devices~\cite{yadav2016observation,hong2017stability,das2019observation,wang2020polar,das2021local,han2022high,khalyavin2020emergent,rusu2022ferroelectric}. In other words, the HfO$_2$ and related materials [{\it e.g.}, (Hf, Zr)O$_2$ and Y-doped HfO$_2$] may be ideal candidates to explore novel electric topological textures. 
Besides, we hope that our work will motivate the discoveries of a sequence of eASEI-based phenomena such as the dipole patterns as counterparts of the mASEI-driven skyrmionic states in two-dimensional magnets~\cite{amoroso2020spontaneous}. This will deepen the current knowledge of electromagnetism in condensed matter systems such as ferroelectrics, magnets and multiferroics. \\

\noindent
\textit{Acknowledgements.--}This work was supported by the National Key Research and Development Program of China (Grant No. 2022YFA1402502) and the National Natural Science Foundation of China (Grant No. 12274174, No. 52288102, and No. 12034009). P.C. and L.B. thank  the Office of Naval Research (ONR) under Grant No. N00014-17-1-2818 and the Vannevar Bush Faculty Fellowship (VBFF) Grant No.
N00014-20-1-2834 from the Department of Defense. L.J.Y. acknowledges the support from the high-performance computing center of Jilin University and the support from the International Center of Future Science, Jilin University. The authors thank Prof. M. Alexe for valuable discussion on the eDMI-related phenomena.\\

%\bibliography{ref}

\end{document}